\documentclass[12pt]{iopart}

\usepackage{cite}

\usepackage{graphicx}

\begin{document}

\title[On numerical solutions of a coupled MKdV system]
{Comment on the numerical solutions of a new coupled MKdV system (2008 Phys.
Scr. 78 045008)}

\author{Francisco M. Fern\'{a}ndez}

\address{INIFTA (UNLP, CCT La Plata-CONICET), Divisi\'on Qu\'imica Te\'orica,
Diag 113 S/N,  Sucursal 4, Casilla de Correo 16,
1900 La Plata, Argentina}\ead{fernande@quimica.unlp.edu.ar}

\maketitle

\begin{abstract}
In this comment we point out some wrong statements in the paper by Inc and
Cavlak, Phys. Scr. 78 (2008) 045008
\end{abstract}

\submitto{\PS}

\pacs{02.30.Jr,02.60.-x,04.20.Jb}

In a recent paper appeared in this journal Inc and Cavlak\cite{IC08} applied
the Adomian decomposition method (ADM) and the variational iteration method
(VIM) to a new coupled modified Korteweg--de Vries (MKdV) system. The
authors state that ``The methods provide the solution in a convergent series
with components that are elegantly computed. The VIM and the decomposition
method avoid the complexity provided by other pure numerical methods''. In
what follows we analyse the methods proposed by the authors and determine if
this claim is true.

Inc and Cavlak\cite{IC08} studied nonlinear partial differential equations
of the form
\begin{equation}
\mathbf{u}_{t}=\mathbf{f}(\mathbf{u},\mathbf{u}_{t},\mathbf{u}_{x},\mathbf{u}%
_{tt},\mathbf{u}_{xx},\mathbf{u}_{tx},\ldots )  \label{eq:dif_eq}
\end{equation}
where $\mathbf{u}$ is a vector of components $u_{1}(x,t),u_{2}(x,t),\ldots
,u_{n}(x,t)$, $\mathbf{f}$ is a vector of nonlinear functions $%
f_{1},f_{2},\ldots ,f_{n}$ and the subscripts $t$ and $x$ indicate
differentiation with respect to these variables. Inc and Cavlak\cite{IC08}
chose a problem with an exact solution that is sufficiently simple to
facilitate the application of both the ADM and VIM. It is the kind of
tailor--made toy problems that are always selected for the application of
such approaches.

Inc and Cavlak\cite{IC08} applied the ADM and VIM in such a way that they
merely obtained the time--power series for the solutions:
\begin{equation}
\mathbf{u}(x,t)=\sum_{j=0}^{\infty }\mathbf{u}_{j}(x)t^{j}
\label{eq:u_t_series}
\end{equation}
To be precise, the ADM yielded the pure Taylor expansion about $t=0$ term by
term and the VIM gave it in a rather mixed way but it is expected that
cancellation of terms in the summation of the contributions would give
exactly the same series. In any way, it is most striking that the authors
had resorted to more or less complicated methods to obtain a time series
that one derives more easily and straightforwardly by simply substituting
Eq.~(\ref{eq:u_t_series}) into equation~(\ref{eq:dif_eq}) and equating the
coefficients of the polynomials in the lhs and rhs. In this way one obtains
a recurrence relation that completely determines the coefficients of the
time series (\ref{eq:u_t_series}) provided that one knows $\mathbf{u}(x,0)=%
\mathbf{u}_{0}(x)$. The authors did not indicate any advantage of those
methods with respect to the well--known Taylor series. Since working harder
to obtain the same results is just a matter of taste we will not discuss
this point any further. We just wanted to call the reader's attention on it.

Inc and Cavlak\cite{IC08} compared their time--power series with the exact
solution for some values of $t$ and $x$ and concluded that ``Numerical
approximations show a high degree of accuracy, and in most cases of $\phi
_{n}$, the $n$--term approximation is accurate for quite low values of $n$.
The proofs of the convergence were investigated by Cherruault and
co-operator'' (and gave some references that are unnecessary for our
purposes as we shall see below). Later they also stated that ``The errors
obtained by using the approximate solution are given by using only two
iterations of the decomposition method. The error is smaller for values of $t
$ close to the initial point $0$. For values of $t$ away from $0$, the error
is decreasing {\it(we believe that the authors meant increasing)}.
However the overall errors can be made even smaller by adding
more iterates. The convergence is rapid.'' Of course the reader will not
doubt that the accuracy of the Taylor expansion of the solutions about $t=0$
will decrease as we move away from the time origin.

The exact solutions to the problem chosen by Inc and Cavlak\cite{IC08} are:
\begin{eqnarray}
u(x,t) &=&1+\frac{1}{2}\tanh \left( x-\frac{11t}{2}\right)   \nonumber \\
v(x,t) &=&1-\frac{1}{4}\tanh \left( x-\frac{11t}{2}\right)   \nonumber \\
z(x,t) &=&2-\tanh \left( x-\frac{11t}{2}\right)   \label{eq:u,v,z}
\end{eqnarray}
Everybody knows that the $\tanh (\theta )$ is singular at $\theta =(2j+1)\pi
i/2$, $j=0,1,\ldots $. From the singular point closest to the origin we
determine that the convergence radius of the time--power series expansion
will be $R(x)=(2/11)\sqrt{x^{2}+\pi ^{2}/4}$. Since this series will not
converge for $t>R(x)$ we conclude that the authors' statements quoted above
cannot be true. Inc and Cavlak\cite{IC08} showed results for $%
x=-15,-10,-5,5,10$ and $t=0.1,0.2,0.3,0.4,0.5$. Notice that the smallest
convergence radius $R(\pm 5)=0.9528972974$ is considerably larger than the
largest $t$--value in the authors' tables. In other words, the pairs of $x,t$
values are conveniently chosen to support the authors' conclusions.
Obviously, the most unfavourable case is $R(0)=\pi /11=0.2855993321$. In our
opinion there is no necessity for a numerical verification of present
arguments. However, we have decided to add a graphical exemplification of
them because of a negative experience with a referee regarding a similar
criticism about a paper in another journal (see below).

Fig.~\ref{fig:MKdV} shows $\tanh (11\,t/2)$ (the relevant term when $x=0$) and
the Taylor series of degree $5$ (the greatest order chosen by Inc and Cavlak%
\cite{IC08}). We clearly appreciate how the accuracy of the series
deteriorates as time approaches $R(0)$. Fig.~\ref{fig:MKdV} also shows that
increasing
the degree of the Taylor series to order $15$ does not do much to improve
this behaviour. Clearly, increasing the order of the ADM or VIM will not
correct the essential limitation of the approaches that in the end produce a
time series. This obvious fact also contradicts the authors' statements
quoted above.

Throughout this comment we have tried to prove two points. First, that the
well--known Taylor--series expansion provides the same kind of results that
Inc and Cavlak\cite{IC08} obtained by the more complicated ADM and VIM.
The reader may decide
if the application of any of those {\it elegant} approximate methods is worth the
extra effort.
Second, the
resulting series are suitable only in a neighbourhood of $t=0$. The reason
is that nonlinear equations spontaneously generate singular points of the
type described above.

In the study of nonlinear systems one is primarily interested in their
overall picture, namely the qualitative and quantitative long--time
behaviour of their solutions. A power series can never provide such
information. It is unlikely that one may be interested in what happens in
the early times of the phenomenon. If that were the case one may try the
Taylor expansion and improve it by means of, for example, Pad\'{e}
approximants. In fact, Pad\'{e} approximants overcome the problem of the
singular points discussed above.

We have raised this kind of criticisms before\cite
{F07,F08b,F08c,F08d,F08e,F08f} but some journals are unwilling to publish
comments on some of the papers they publish. This journal seems to exhibit a
different policy in this regard\cite{Fr08}.

\begin{figure}[H]
\begin{center}
\includegraphics[width=9cm]{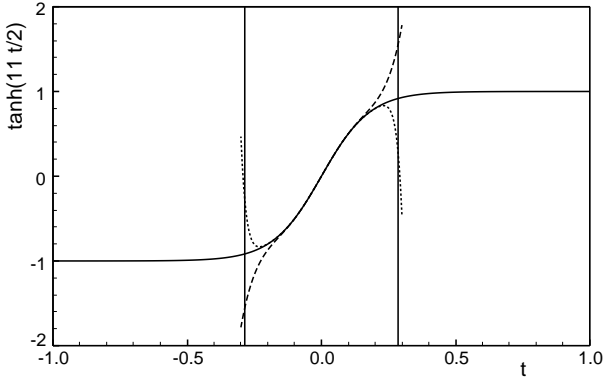}
\end{center}
\caption{Exact $\tanh(11\,t/2)$ (solid line) and its Taylor expansions of
degree $5$ (dashed line) and $15$ (dotted line). The vertical lines bound the
convergence interval.}
\label{fig:MKdV}
\end{figure}

\end{document}